\def\papertitle{Estimates of the Reconstruction Error in Partially Redressed Warped Frames Expansions}
\def\paperauthorA{Thomas Mejstrik}
\def\paperauthorB{Gianpaolo Evangelista}

\documentclass[twoside,a4paper]{article}
\usepackage{dafx_16}
\usepackage{amsmath,amssymb,amsfonts,amsthm}
\usepackage{euscript}
\usepackage[latin1]{inputenc}
\usepackage[T1]{fontenc}
\usepackage{ifpdf}

\usepackage[english]{babel}
\usepackage{caption}
\usepackage{subfig, color}

\usepackage{booktabs}
\usepackage[dvipsnames,table]{xcolor}

\usepackage{nicefrac}
\usepackage{paralist}

\ninept

\newif\ifpdf
\ifx\pdfoutput\relax
\else
   \ifcase\pdfoutput
      \pdffalse
   \else
      \pdftrue
\fi

\ifpdf 
  \usepackage[pdftex,
    pdftitle={\papertitle},
    pdfauthor={\paperauthorA, \paperauthorB},
    colorlinks=false, 
    bookmarksnumbered, 
    pdfstartview=XYZ 
  ]{hyperref}
  \usepackage[pdftex]{graphicx}
  \usepackage[figure,table]{hypcap}
\else 
  \usepackage[dvips]{epsfig,graphicx}
  \usepackage[dvips,
    colorlinks=false, 
    bookmarksnumbered, 
    pdfstartview=XYZ 
  ]{hyperref}
  \usepackage[figure,table]{hypcap}
\fi

\title{\papertitle}

\twoaffiliations{
	\paperauthorA \, \sthanks{Part of the research contained in this work was performed by this author while he was a Master student at MDW, Wien, under the guidance of Prof. Gianpaolo Evangelista}}
{\href{https://mathematik.univie.ac.at/en/main/}{Department of Mathematics} \\ University of Vienna \\ Vienna, Austria\\
	{\tt \href{mailto:tommsch@gmx.at}{tommsch@gmx.at}}
}
{\paperauthorB \,}
{\href{http://mdw.ac.at}{Institute for Composition and Electroacoustics,} \\ MDW, University of Music and Performing Arts\\ Vienna, Austria\\
	{\tt \href{mailto:evangelista@mdw.ac.at}{evangelista@mdw.ac.at}}
}

\newcommand{\RR}{\mathbb{R}}
\newcommand{\NN}{\mathbb{N}}

\newcommand{\abs}[1]{\left|#1\right|}

\newcommand{\WL}{\text{T}}
\newcommand{\avg}{\text{avg}}
\newcommand{\cut}{\text{cut}}
\newcommand{\BW}{\text{B}}
\newcommand{\SR}{\text{SR}}

\newcommand{\ttphi}{\tilde{\tilde{\varphi}}}

\newcommand{\sinc}{\text{sinc}}
\newcommand{\innerproduct}[2]{\langle #1,#2\rangle}
\newcommand{\set}[2][]{\lbrace #2 \rbrace_{#1}}

\newcommand{\cred}[1]{{\color{BrickRed}#1}}
\newcommand{\cgrn}[1]{{\color{PineGreen}#1}}

\usepackage[usestackEOL]{stackengine}

\begin{document}
\ifpdf 
  \DeclareGraphicsExtensions{.png,.jpg,.pdf}
\else  
  \DeclareGraphicsExtensions{.eps}
\fi

\maketitle
\begin{abstract}
	In recent work, redressed warped frames have been introduced for the analysis and synthesis of audio signals with non-uniform frequency and time resolutions. In these frames, the allocation of frequency bands or time intervals of the elements of the representation can be uniquely described by means of a warping map. Inverse warping applied after time-frequency sampling provides the key to reduce or eliminate dispersion of the warped frame elements in the conjugate variable, making it possible, e.g., to construct frequency warped frames with synchronous time alignment through frequency. 
	The redressing procedure is however exact only when the analysis and synthesis windows have compact support in the domain where warping is applied. This implies that frequency warped frames cannot have compact support in the time domain. This property is undesirable when online computation is required. Approximations in which the time support is finite are however possible, which lead to small reconstruction errors. In this paper we study the approximation error for compactly supported frequency warped analysis-synthesis elements, providing a few examples and case studies.
\end{abstract}

\section{Introduction}\label{sec:introduction}
The availability of configurable time-frequency schemes is an asset for sound analysis and synthesis, where the elements of the representation can efficiently capture features of the signal. These include features of 
interpretation: e.g. glissandi and vibrati; 
human perception: e.g. non-uniform
frequency sensitivity of the cochlea; 
music theory: e.g. scales (pentatonic, 12-tone, Indian scales with unequally spaced tones, etc.); physical effects: e.g. non-harmonic overtones in the low register of the piano or of percussion instruments and so forth. When used in the context of Music Information Retrieval, the adaptation of the representation to music scales is bound to improve. e.g. for instrument-, note-, chord- and overtone detection and recognition.

Traditionally, two extreme cases have been considered: Gabor expansions featuring uniform time and frequency resolutions and orthogonal wavelet expansions and frames featuring octave band allocation and constant uncertainty of the representation elements. In previous work \cite{dohove11,badohojave11,EvaDoeMat12, HolWies14, Twaroch98, Baraniuk93, Braccini74}, 
generalised Gabor frames have been constructed which allow for non-uniform time-frequency schemes with perfect reconstruction. In \cite{EvaDoeMat12} the allocation of generalised Gabor atoms is specified according to a frequency or time warping map. In \cite{Eva13} the STFT redressing method is introduced, which, with the use of additional warping in time-frequency, shows under which conditions one can have generalised Gabor frames. These conditions stem from the interaction of sampling in time-frequency and frequency or time warping operators, which allow to incorporate the results in \cite{EvaDoeMat12} in a more general context. It is shown that arbitrary allocation of the atoms is exactly possible in the so called \emph{painless case}, i.e. in the case of finite time support of the windows for arbitrary time interval allocation and of finite frequency support of the windows for arbitrary frequency band allocation.
Non-uniform frequency analysis by means of warping was introduced in~\cite{Braccini74}. Non-uniform Gabor frames with constant-Q were previously introduced in~\cite{dohove11} , based on the theory developed in~\cite{badohojave11}, where an ad hoc procedure was employed for their construction. In~\cite{EvaDoeMat12} we provided an alternative
general method for their construction, using warping. In~\cite{Eva13} the redressing method was introduced and applied to the general construction
of non-stationary Gabor frames. In~\cite{Eva14} the general theory was revisited with the real-time computational aspects in mind. There, the first approximate schemes were introduced without extensive testing. 

Since online computation of the generalised Gabor analysis-synthesis is only possible with finite duration windows, the arbitrary frequency band allocation does not lead to an exact solution in applications that require real-time, while the arbitrary time interval allocation presents little or no problem. In \cite{Eva14} approximations leading to nearly exact representations were introduced. 

In this paper we expand on the results in \cite{Eva14} and provide a study of the approximation error on a wide class of signals when finite duration windows are required in arbitrary frequency band allocation.

The paper is organised as follows. In Section \ref{sec:redressed_gabor} we review the concept of applying time and frequency warping to time-frequency representations, together with the redressing method, which involves a further warping operation in the time-frequency domain to reduce or eliminate dispersion. In Section \ref{sec:online} we introduce approximations suitable for the online computation of redressed frame expansions. In Section \ref{sec:error_estimates} the results of numerical experiments are shown, which provide estimations of the approximation error. In Section \ref{sec:conclusions} we draw our conclusions.

\section{Redressed Warped Gabor Frames}\label{sec:redressed_gabor}
In this section we review the concepts leading to the definition of redressing in the context of frequency warped time-frequency representations. First we review the basic notions of STFT (Short-Time Fourier Transform) and Gabor frames. Then we move on to the definition of warped frames and then to the redressing procedure.

\subsection{STFT and Gabor Frames}
Gabor expansions can be considered as a form of sampling and exact reconstruction of the STFT. As is well-known, given a window $h$ and defining the time-shift operator ${\mathbf{T}}_\tau  s(t) = s(t - \tau )$ and the modulation operator ${\mathbf{M}}_\nu  s(t) = e^{j2\pi\nu t} s(t)$, the STFT is obtained by applying the operator $\mathcal{S}$ to the signal $s$:
\begin{equation}
\begin{gathered}
\left[ {\mathcal{S} s} \right](\tau ,\nu ) = \left\langle {s,h_{\tau ,\nu } } \right\rangle  = \left\langle {s,{\mathbf{T}}_\tau  {\mathbf{M}}_\nu  h} \right\rangle  = \\ \int_{ - \infty }^{ + \infty } s (t)\overline{h_{0,0}(t - \tau )}e^{ - j2\pi\nu (t - \tau )} dt,
\label{STFT}
\end{gathered}
\end{equation}
where the overbar denotes complex conjugation. 

A {\em  Gabor system} is generated by the kernel of $\mathcal{S}$ by sampling the time-frequency plane $(\tau ,\nu)$:
\begin{equation}
{\mathcal G}(h,a,b) = \{\mathbf{T}_{na} \mathbf{M}_{qb}  h : n,q \in \mathbb{Z} \}
\end{equation}
where $a, b > 0$ are sampling parameters. 

The scalar products of the signal with the members of the Gabor system
\begin{equation}
\langle s,\mathbf{T}_{na} \mathbf{M}_{qb}  h \rangle, \qquad n,q \in \mathbb{Z}
\label{STFT_samples}
\end{equation}
provide evaluations of the STFT (\ref{STFT}) of a signal $s$ with window $h$ at the time-frequency grid of points $( na, qb )$, with  $n,q \in \mathbb{Z} $. The question whether the signal $s$ can be reconstructed from these evaluations can be addressed by introducing the concept of frame. 

A sequence of functions $\left\{ \psi_l\right\}_{l\in I}$ in the Hilbert space $\mathcal{H}$ is called a frame if there exist both positive constant lower and upper bounds $A$ and $B$, respectively, such that
\begin{equation} 
A\|s\|^2\leq\sum_{l\in I}|\langle s,\psi_l\rangle|^2 \leq
B\|s\|^2\quad\forall s \in \mathcal{H},
\vspace{-2mm}
\label{framedef}
\end{equation}
where $\|s\|^2=\langle s,s\rangle$ is the norm square or total energy of the signal. Frames generate signal expansions, i.e., the signal can be perfectly reconstructed from its projections over the frame.

A Gabor system that is a frame is called a \emph{Gabor frame}. In this case, the signal can be reconstructed from the corresponding samples of the STFT (\ref{STFT_samples}). While not unique, reconstruction can be achieved with the help of a dual frame, which in turn is a Gabor frame generated by a dual window ${\tilde h}$. Perfect reconstruction essentially depends on the choice of the window and the sampling grid. One can show that there exist no Gabor frames when $ab>1$. See~\cite{GROE2001} for more informations about Gabor frames.

\subsection{Warped STFT and Gabor Frames}
The warped STFT can be obtained by warping the signal prior to applying the STFT operator. In this paper we focus on pure frequency warping.

A frequency warping operator $\mathbf{W}_{\tilde\theta}$ is completely characterised by a function composition operator $\mathbf{W}_{\theta}$, such that  ${{\mathbf{W}}_{\theta} x} = x \circ \theta$, in the frequency domain:
\begin{equation}
\mathbf{W}_{\tilde\theta} = \mathcal{F}^{-1} \mathbf{W}_{\theta} \mathcal{F},
\end{equation}
where $\mathcal{F}$ is the Fourier transform operator. The function $\theta$ is the frequency warping map, which transforms the Fourier transform $\hat s=\mathcal{F}s$ of a signal $s$ into the Fourier transform $\hat s_{fw}=\mathcal{F}s_{fw}$ of another signal $s_{fw}$. We affix the $\tilde{}$ symbol over the map $\theta$ as a reminder that the map operates in the frequency domain. 

If the warping map is one-to-one and almost everywhere differentiable then a unitary form $\mathbf{U}_{\tilde \theta}$ of the warping operator can be defined by the following frequency domain action
\begin{equation}
\hat s_{fw}(\nu)=\left[\widehat {\mathbf{U}_{\tilde \theta} s} \right](\nu)=\sqrt {\left|\tfrac{{d\theta }}{{d\nu }}\right|}\hat s(\theta (\nu )),
\label{unitary_warping_op}
\end{equation}
where $\nu$ denotes frequency. We assume henceforth that all warping maps are almost everywhere increasing so that the magnitude sign can be dropped from the derivative under the square root. 

Given a frequency warping operator $\mathbf{W}_{\tilde\theta}$, the warped STFT is defined through the operator $\mathcal{S}_{\tilde\theta}$ as follows
\begin{equation}
\begin{gathered}
\left[ \mathcal{S}_{\tilde\theta} s \right](\tau ,\nu ) = \left[ \mathcal{S} {\mathbf{W}}_{\tilde\theta} s \right](\tau ,\nu ) =\\ \left\langle {{\mathbf{W}}_{\tilde\theta} s,h_{\tau ,\nu } } \right\rangle = \left\langle {s,{\mathbf{W}}^\dagger_{\tilde\theta} h_{\tau ,\nu } } \right\rangle, 
\end{gathered}
\label{WarpedSTFT}
\end{equation}
which is indeed a warped version of (\ref{STFT}), where $\mathbf{W}^\dagger_{\tilde\theta}$ is the adjoint of the warping operator. If the warping operator is unitary then we have $\mathbf{W}^\dagger_{\tilde\theta}=\mathbf{W}_{\tilde\theta}^{-1}=\mathbf{W}_{\tilde\theta^{-1}}$. In that case, warping the signal prior to STFT is perfectly equivalent to perform STFT analysis with inversely frequency warped windows. The warped STFT is unitarily equivalent to the STFT so that a number of properties concerning conditioning and reconstruction hold \cite{Bar95}.

The Fourier transforms of the frequency warped STFT analysis elements are
\begin{equation}
\begin{gathered}
\hat{ \tilde h}_{\tau ,\nu } (f ) = \left[ {\widehat{\mathbf{W}_{\tilde\theta^{ - 1}}  h_{\tau ,\nu } }} \right](f ) =\\ \sqrt {\tfrac{{d\theta ^{ - 1} }}
	{{df }}} \hat h (\theta ^{ - 1} (f ) - \nu )e^{ - j2\pi\theta ^{ - 1} (f )\tau }, 
\label{W_analysis_elements}
\end{gathered}
\end{equation}
i.e., the warped STFT analysis elements are obtained from frequency warped modulated windows centred at frequencies $f=\theta(\nu)$. The windows are time-shifted with dispersive delay, where the group delay is 
$ \tau\tfrac{{d\theta ^{ - 1} }}{{df }} $.

Frequency warping generally disrupts the time organisation of signals due to the fact that the time-shift operator ${\mathbf{T}}_\tau$
does not commute with the frequency warping operator \cite{Eva14}.

From (\ref{framedef}) it is easy to see that any unitary operation, in particular unitary warping, on a frame results in a new frame with the same frame bounds $A$ and $B$ \cite{Bar95}. Since the atoms are not generated by shifting and modulating a single window function, the resulting frames are not necessarily of the Gabor type. However, warping prior to conventional Gabor analysis and unwarping after Gabor synthesis always leads to perfect reconstruction.

Starting from a Gabor frame (analysis) $\left\{ {{\varphi}_{n,q} } \right\}_{q,n \in {\mathbb{Z}}}$ and dual frame (synthesis) $\left\{ {{\gamma}_{n,q} } \right\}_{n,q \in {\mathbb{Z}}}$:
\begin{equation}
\begin{gathered}
{\varphi}_{n,q}  =  \mathbf{T}_{na } \mathbf{M}_{qb}  h \hfill \\
{\gamma}_{n,q}  = \mathbf{T}_{na } \mathbf{M}_{qb}  g, \hfill \\
\end{gathered}
\label{gabor_frame}
\end{equation}
where $h$ and $g$ are dual windows, warped frames can be generated, following (\ref{W_analysis_elements}), by unwarping the analysis and synthesis frames. In the case of non-unitary warping, a frequency domain scaling operation is necessary in order to reconstruct the original signal. For the case of unitary warping we simply have: 
\begin{equation}
\begin{gathered}
\tilde{\varphi}_{n,q}  = \mathbf{U}_{\tilde {\theta}}^\dag {\varphi}_{n,q} = \mathbf{U}_{\tilde {\theta}^{-1}} \mathbf{T}_{na } \mathbf{M}_{qb}  h \hfill \\
\tilde{\gamma}_{n,q}  = \mathbf{U}_{\tilde {\theta}}^\dag {\gamma}_{n,q} = \mathbf{U}_{\tilde {\theta}^{-1}} \mathbf{T}_{na } \mathbf{M}_{qb}  g ,\hfill \\
\end{gathered}
\label{warped_gabor_frame}
\end{equation}
where $\left\{ {\tilde{\varphi}_{n,q} } \right\}_{q,n \in {\mathbb{Z}}}$ is the frequency warped analysis frame and $\left\{ {\tilde{\gamma}_{n,q} } \right\}_{n,q \in {\mathbb{Z}}}$ is the dual warped frame for the synthesis. With these definitions, one obtains the signal expansion
\begin{equation}
s = \sum\limits_{n,q \in Z} {\left\langle {s,\tilde \varphi _{n,q} } \right\rangle } \tilde \gamma _{n,q} .
\end{equation}

Warped Gabor frames suffer from the same problem as the warped STFT. Indeed the Fourier transforms of the warped Gabor frame elements bear frequency dispersive delays so that dispersive time samples are produced by the direct application of the frequency warped frame analysis. 

\subsection{Redressing Methods}\label{sec:redressing}
As shown in \cite{Eva13,Eva14}, the dispersive delays intrinsic to the warped STFT can be redressed, i.e. made into constant delays in each analysis band if frequency unwarping is performed in the transformed time domain, i.e. with respect to time shift. In other words, instead of (\ref{WarpedSTFT}) we consider the similarity transformation ${{\mathbf{W}_{\tilde\theta}^\dagger \mathcal{S} \mathbf{W}_{\tilde \theta}^{ } }}$ on the STFT operator, which is time-shift covariant. In fact, one has:
\begin{equation}
\left[ {\widehat{\mathbf{W}_{\tilde\theta^{ - 1}} \mathcal{S} \mathbf{W}_{\tilde\theta} s}} \right](f ,\nu ) =  \overline {\hat h_{0,0 } (\theta^{-1}(f)-\nu )} \hat s(f ),
\label{redressed_WSTFT}
\end{equation}
which is in the form of a time-invariant filtering operation, corresponding to convolution in time domain. The filters are frequency warped versions of the modulated windows in the traditional STFT. The Fourier transform of the redressed analysis elements are
\begin{equation}
\hat{ \tilde{\tilde h}}_{\tau ,\nu } (f ) = \left[ {\widehat{\mathbf{T}_\tau \mathbf{W}_{\tilde\theta}  h_{0 ,\nu } }} \right](f ) =  \hat h_{0,0} (\theta ^{ - 1} (f ) - \nu )e^{ - j2\pi f \tau },
\label{redressed_analysis_elements}
\end{equation}
which shows that the dispersive delays in the analysis elements (\ref{W_analysis_elements}) are brought back to non-dispersive delays.

In redressing warped Gabor frames one faces a further difficulty due to time-frequency sampling. In this case, inverse frequency warping can only be applied to sequences (with the respect to the time shift index) and may not perfectly reverse the dispersive effect of the original map on delays.

Unitary frequency warping in discrete time can be realised with the help of an orthonormal basis of $\ell^2(\mathbb{Z})$ constructed from an almost everywhere differentiable warping map $\vartheta$ that is one-to-one and onto $[-\tfrac{1}{2},+\tfrac{1}{2}[$, as follows:
\begin{equation}
\mu _m (n) = \int_{ - \tfrac{1}{2}}^{ + \tfrac{1}{2} } {\sqrt {\tfrac{{d\vartheta }}
		{{d\nu }}} } e^{j2\pi(n\vartheta (\nu )  - m\nu)} d\nu ,
\end{equation}
where $n,m \in \mathbb{Z}$ (see \cite{Bro65,Kno01,Eva01a,EvaCav98a,EvaCav98b}). The map $\vartheta$ can be extended over the entire real axis as congruent modulo $1$ to a $1$-periodic function.

Given any sequence $\{x(n)\}$ in $\ell^2(\mathbb{Z})$, the action of the discrete-time unitary warping operator $\mathbf{D}_{{\tilde\vartheta}}$ is defined as follows:
\begin{equation}
\tilde{x}(m) = \left[ \mathbf{D}_{{\tilde\vartheta}}x\right](m)= \left\langle {x,\mu _m } \right\rangle _{\ell^2(\mathbb{Z})}.
\end{equation}
In fact, the sequence $\{\tilde{x}(m)\}$ in $\ell^2(\mathbb{Z})$ satisfies
\begin{equation}
\hat{\tilde{x}}(\nu ) = \sqrt {\tfrac{{d\vartheta }}
	{{d\nu }}} \hat {x}(\vartheta (\nu )),
\end{equation}
where the $\hat{}$ symbol, when applied to sequences, denotes discrete-time Fourier transform. The sequences $\overline{\eta _m (n)}$ define the nucleus of the inverse unitary frequency warping $\ell^2(\mathbb{Z})$ operator $\mathbf{D}_{{\tilde\vartheta}^{-1}}^{ }=\mathbf{D}_{{\tilde\vartheta}}^\dag$. where $\eta _m (n)={\mu _n (m)}$.

In order to limit or eliminate time dispersion in the frequency warped Gabor expansion, the discrete-time frequency warping operator $\mathbf{D}_{{\tilde\vartheta}^{-1}}$ is applied to the time sequence of expansion coefficients over the warped Gabor frames. Since the operator is applied only on the time index, for generality, one can include dependency of the map and of the sequences $\eta_n$ on the frequency index $q$. The process can be equivalently described by defining the redressed frequency warped Gabor analysis and synthesis frames as follows:
\begin{equation}
\begin{gathered}
\tilde{\tilde\varphi}_{n,q}  = \mathbf{D}_{{\tilde{\vartheta}_q}^{-1}}\tilde\varphi _{\bullet,q} =\sum\limits_m { {\eta _{n,q} (m)} \tilde\varphi _{m,q} } \hfill \\
\tilde{\tilde\gamma}_{n,q}  = \mathbf{D}_{{\tilde{\vartheta}_q}^{-1}}\tilde\gamma _{\bullet,q} =\sum\limits_m { {\eta _{n,q} (m)} \tilde\gamma _{m,q} } , \hfill
\end{gathered}
\label{red_frames}
\end{equation}
obtaining:
\begin{equation}
s = \sum\limits_{n,q \in \mathbb(Z)} {\left\langle {s,\tilde {\tilde{ \varphi}} _{n,q} } \right\rangle } \tilde {\tilde{ \gamma}} _{n,q}.
\end{equation}

One can show \cite{Eva14} that the Fourier transforms of the redressed frame are
\begin{equation}
\hat {\tilde {\tilde {\varphi}}} _{n,q} (f ) = A(f) \hat h(\theta ^{ - 1} (f ) - qb)e^{ - j2\pi n\vartheta _q (a\theta ^{ - 1} (f ))} ,
\label{FT_redressed_frame}
\end{equation}
where
\begin{equation}
A(f)=\sqrt {\tfrac{{d\theta ^{ - 1} }}
	{{df }}} \left. {\sqrt {\tfrac{{d\vartheta _q }}
		{{d\nu }}} } \right|_{\nu  = a\theta ^{ - 1} (f )}.
\label{redressed_dispersion_scaling_fact}
\end{equation}
Hence, dispersion is completely eliminated if 
\begin{equation}
\vartheta _q (a\theta ^{ - 1} (f ))=d_q f
\label{delay_cond}
\end{equation}
for any $f \in \mathbb{R}$, where $d_q$ are positive constants controlling the time scale in each frequency band. In this case, the Fourier transforms of the redressed frame elements become:
\begin{equation}\label{equ:ttphi}
\hat {\tilde {\tilde {\varphi}}} _{n,q} (f ) = \sqrt {\tfrac{{d_q }}
	{a}} \hat{h}(\theta ^{ - 1} (f ) - qb)e^{ - j2\pi nd_q f } .
\end{equation}
When all $d_q$ are identical all the time samples are aligned to a uniform time scale throughout frequencies. If the $d_q$ are distinct, time realignment when displaying the non-uniform spectrogram is a simple matter of different time base or time scale for each frequency band. 

Unfortunately, due to the discrete nature of the redressing warping operation, each map $\vartheta_q$ is constrained to be congruent modulo $1$ to a $1$-periodic function, while the global warping map $\theta$ can be arbitrarily selected. Moreover, the functions $\vartheta_q$ must be one-to-one in each unit interval, therefore they can have at most an increment of $1$ there. 

\begin{figure}[t]
	\centering
	\includegraphics[width=0.95\columnwidth]{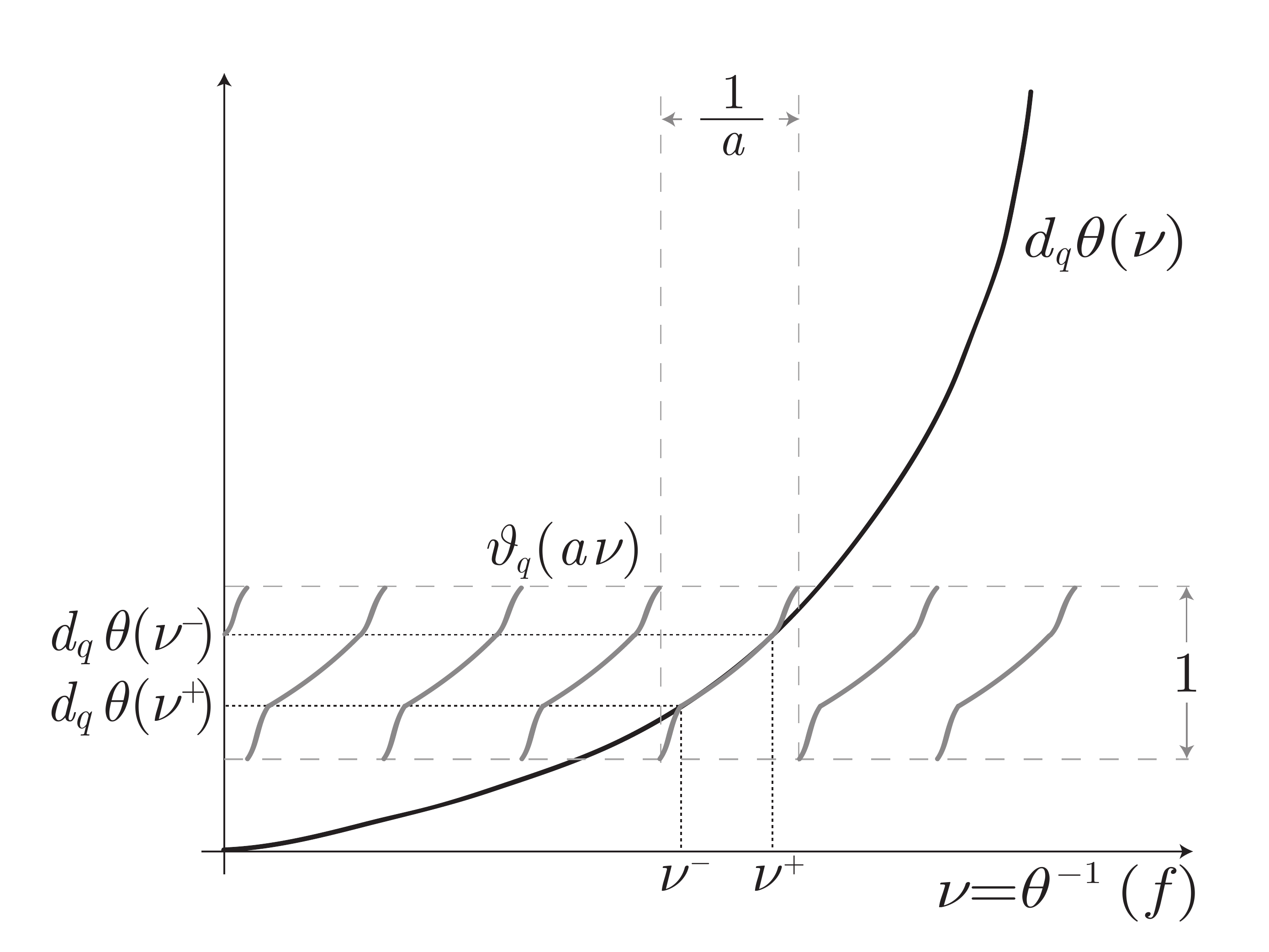}
	\caption{Locally eliminating dispersion by means of discrete-time frequency warping. Black line: curve derived from the original map $\theta$ by amplitude scaling. Gray line: discrete-time frequency warping characteristics for local phase linearisation.
		\label{fig:DelayElim}}
\end{figure}

In Fig.~\ref{fig:DelayElim} we illustrate the phase linearisation problem. There, the black curve is the amplitude scaled warping map $d_q \theta(\nu)$ and the grey curve represents the map $\vartheta_q(a\nu)$, which is $1/a$-periodic. Both maps are plotted in the abscissa $\nu=\theta^{-1}(f)$. By amplitude scaling the warping map $\theta$ one can allow the values of the map to lie in the range of the discrete-time warping map $\vartheta_q$. The amplitude scaling factors happen to be the new time sampling intervals $d_q$ of the redressed warped Gabor expansion. 

In the ``painless'' case, which was hand picked in  \cite{EvaDoeMat12}, the window $h$ is chosen to have compact support in the frequency domain. Through equation (\ref{FT_redressed_frame}) and condition (\ref{delay_cond}), the redressing method shows that, given any continuous and almost anywhere differentiable and increasing warping map, only in the painless case one can exactly eliminate the dispersive delays with the help of (\ref{red_frames}). In fact, in this case linearisation of the phase is only required within a finite frequency range given by the frequency support of the frame elements \cite{Eva13,Eva14}, which is compatible with the periodicity constraint of the redressing maps $\vartheta_q$. 

In the general case, a perfect time realignment of the components is not guaranteed. Notice, however, that, by construction, the redressed warped Gabor systems are guaranteed to be frames for any choice of the maps $\vartheta_q$ satisfying the stated periodicity conditions, even when the phase is not completely linearised. Locally, within a certain band of the warped modulated windows it is possible to linearise the phase of the complex exponentials in (\ref{FT_redressed_frame}). In the sequel we will refer to this band as the {\it essential bandwidth} since, hopefully, the magnitude of the Fourier transform of the window is negligible outside it, at least as a design goal. 

Unfortunately, in general both the painless case and the partially redressed cases lead to infinite duration windows, which are undesirable when online computation is required. In the next  we are going to propose some approximations and study the reconstruction error also through numerical experiments.  

\section{Real-time computation of the Warped Gabor Expansion}\label{sec:online}
For real-time computation one needs to make assumptions on the signal, as well on the window functions $h$ in order to keep the computational load as low as possible. 

By requiring the window $h$ to be real-valued and $\theta$ to be odd, we
obtain $\hat{\ttphi}_{n,-q}(-f)$ = $\overline{\hat{\ttphi}_{n,q}(f)}$. If, additionally the input signal is real-valued, then the  coefficients $c_{n,q}=\innerproduct{s}{\ttphi_{n,q}}$ fulfil
$c_{n,-q}=\overline{c_{n,q}}$ and thus we only need to compute about half the coefficients
and frame elements. By enforcing shift-invariance of the warped frame elements (i.e. $\ttphi_{n,q}(t)=\ttphi_{0,q}(t-na)$), which we must do since otherwise the pre-computation of the frame elements and hence the real-time computation of the expansion would be impossible, it is sufficient to only store $\ttphi_{0,q}$ for non-negative values of q.
It is advisable that the warping map is selected as a continuously differentiable function, since the error in the resynthesised signal in the frequency domain at the points of discontinuity of $\theta'$ is very high. 
A strictly positive derivative helps the atoms not get too stretched in the time domain, which is undesirable in real time applications. To avoid aliasing we further require that $\theta(qb)=\SR/2$ and $\theta''(qb)=0$ for some $q\in\NN$, where $\SR$ denotes the sampling rate. This ensures that the high frequencies are smoothly mapped back to the negative frequencies and avoids extra aliasing introduced by the approximation.
By requiring the Fourier transform of the window $h$ to have an analytic expression, we can avoid numerical errors in the computation of the warped windows.
We further only worked with windows which are dual to itself, i.e. $\varphi_{n,q}=\gamma_{n,q}$ which is actually not a necessary restriction.

\subsection{Implementation Details}
We implemented the approximate warped redressed Gabor analysis and synthesis as C externals interfaced to Pure-Data (32\,bit and 64\,bit). It runs both under Windows and Linux and can use multiple cores. The warped windows are computed using equation~\eqref{equ:ttphi} and applying an IDFT of that data.
The inner products $c_{n,q}$ are directly computed with a loop. Also for the synthesizing part, the sum $\sum_{n,q} c_{n,q}\ttphi_{n,q}$ is directly computed. It turned out that this is sufficiently fast for real-time computation and whence we made no use of fast convolution algorithm.
Since the data rate of the analysis part is not uniform in time, PD's signal connections are not suitable to transfer the coefficients. Therefore we use the signal connections to transfer pointers rather than signals.

\subsection{Interface Details}
For simplicity we require that the essential bandwidth $\BW$ is a multiple $K\in\NN$ of the frequency shift parameter $b$, i.e. $\BW=Kb$. Then, in order to obtain a frame the following conditions must be fulfilled as can be seen easily in Figure~\ref{fig:DelayElim}~\cite[eq. (34) to (36)]{Eva14}:
\begin{equation}
\begin{aligned}\label{equ:abR}
abK\leq 1\\
d_q \BW_q\leq 1 
\end{aligned}
\end{equation}
where $\BW_q$ is the ess. bandwidth of the warped modulated window.
\begin{equation}
\BW_q= \theta\left(qb+\frac{\BW}{2}\right)-\theta\left(qb-\frac{\BW}{2}\right)
\end{equation} 

In the case of an exponential increasing warping map, it makes sense to set the frequency shifting parameter in a way that adjacent notes fall away from the windows main lobe in the frequency domain (If there is such a main lobe, as in the case of the raised cosine window). This on the other hand determines the window length $\WL_h$, a minimal value for $R$ and the time shift parameter 
\begin{equation}
a=\frac{T_h}{R}
\end{equation}
with $R\geq K$ due to equation~\eqref{equ:abR}. We set $R=K$ for simplicity.
Equations~\eqref{equ:abR} are fulfilled by setting 
\begin{equation}
\begin{aligned}
b&=\frac{1}{aRC_b}&&
d_q &\simeq \frac{1}{\BW_q C_d}
\end{aligned} 
\end{equation}
where $C_b,C_d$ can be chosen inside the PD-patch. $C_b$ together with $R$ controls the oversampling, where $C_d$ controls the bandwidth in which the phase is linearised and also influences the oversampling. The $d_q$ are different for each band which results in a non-uniform data rate for each band.
 We remark that the numbers $d_q$ have to be chosen, such that $d_q/\SR \in\NN$, where $\SR$ is the sampling rate. 

The number of bands $q_{\sup}$ we need for a given $\SR$ is 
\begin{align}
q_{\sup}=\frac{\theta^{-1}(SR/2)}{b}.
\end{align}
The assumptions on the warping map ensure that this is a natural number.

Due to the warping, the supports of the windows are in general unbounded. Thus we compute the windows with a zero padded array of length of $\WL_c$ ($c$ for compute), which is defined as 
\begin{align}
\WL_c=\frac{\WL_h}{\theta'_{\inf}}\,C_{\WL_c}
\end{align} 
where $C_{\WL_c}\geq 1$ can be chosen  inside the PD-patch and $\theta'_{\inf}=\inf_{f\in\RR}\theta'(f)$. 
Due to the same reason it is indispensable to cut the windows after warping. To define a sensible atom length $T_q$ after warping, we set the atoms to zero after the point from which they were smaller than  $\abs{\ttphi_q(t)}<1/C_{\cut}\max_t\abs{\ttphi_q(t)}$, with $C_{\cut}$ a constant which can be chosen inside the PD-patch. Afterwards the windows are truncated accordingly and with respect to the parameter $\WL_{\max}$, which defines the maximal desired window length. It is important that the windows are cut after aligning them to the time origin.
The second approach suggested in~\cite{Eva14} to obtain windows with finite length by computing only a linearised version of the warped windows, leads to very bad reconstruction.

\begin{figure}[t]
	\centering
	\includegraphics[scale=0.5]{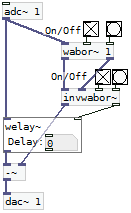}
	\caption{This is a sample patch for the use of our Pure-Data Implementation. \emph{wabor\~} and \emph{invwabor\~} are for analysing and synthesising. They are connected with two signal-paths.  \emph{invwabor\~} holds at the right outlet the delay in samples due to the analysis-synthesis procedure. \emph{welay\~} is a simple delay line. If perfect reconstruction is achieved, the output at \emph{dac\~} is zero. With the toggles the externals can be switched on and off. The bangs are for outputting the parameters used.
	\label{fig:sample_patch}}
\end{figure}

\begin{figure}[t]
	\centering

	\label{fig:parameters}
	\begin{tabular}{  l | l  r | r  r | r  r  }	
	Abbrev. & Explanation\\
	\toprule
	$\SR$	& Sampling Rate\\
	$\BW_q$ & Essential bandwidth of the $q^{th}$ window\\
	$\WL_c$, $C_{\WL_c}$ & Length of the window used in pre-computation\\
	$\WL_h$ & Length of the original window\\
	$\WL_q$ & Length of the q$^{th}$ window\\
	$\WL_{\max}$ & Maximal window length used for \\
&	real time computation\\
	$C_{\cut}$ & Parameter controlling the truncation of the win-\\
&	dows in the time domain after precomputing\\
	$b$, $C_b$ & Frequency shift parameter\\
	$d$, $C_d$ & Time shift parameter of the warped windows\\
	$q_{\sup}$ & Number of bands\\
	\end{tabular}
		\caption{Summary of the variables used in the PD implementation. The parameters $C_X$ control their  variable  $X$.}
\end{figure}			

\subsection{Computational Costs} \label{sec:computational_costs}
\subsubsection{Analysis}

A rough estimation yields the following. Let $\WL_q$ denote the length of the $q^{th}$-window . It will be clear from the context if $\WL_q$ denotes seconds or samples. To compute the inner product of that window with the signal one needs $2 \WL_q$ many real multiplications and summations. This has to be done every $d_q$ samples. Hence, per sample we have $4 \WL_q/d_q$ many floating point operations (flops) in average. If the essential bandwidth is not too large, then $\WL_q$ is proportional to $\WL_0/\theta'(qb)$ and
$d_q$ is proportional to $1/(Kb\theta'(qb))$ since,
linearising $\theta$ around $qb$ yields $\theta(qb+\frac{Kb}{2})\simeq \theta(qb)+\theta'(qb)\frac{Kb}{2}$ and hence
\begin{align}
\BW_q =\theta\left(qb+\frac{Kb}{2}\right)-\theta\left(qb-\frac{Kb}{2}\right)\simeq \theta'(qb)Kb \nonumber\\
\Rightarrow d_q =\frac{1}{\BW_q} \simeq \frac{1}{\theta'(qb)Kb}
\end{align}

Summing up over all windows we get the average number of operations per sample $N_{\avg}$:
\begin{align}\label{equ:navg}
N_{\avg}=\sum_q \frac{4\WL_q}{d_q} 
\sim \sum_q \frac{4\WL_0}{\theta'(qb)} \frac{Kb\theta'(qb)}{1}
\simeq 4K\theta^{-1}(\SR/2)\WL_0
\end{align}
Where $q_{\sup}$, defined above, denotes the number of bands and depends on $\theta^{-1}$.
That means the complexity of the analysis part is proportional to $\WL_0$, $K$ and $b$ and  $q_{\sup} \sim \theta^{-1}(SR)$.

If there is a lower bound for the $d_q$'s, one can choose identical numbers for all bands. However, this increases the computational load tremendously, making real time computation impossible.

The above is an estimation of the average computational cost. In the worst case all inner products of the windows with the signal have to be computed starting at one frame. The number of operations for that frame is
\begin{align}
\sum_q 4\WL_q\simeq4\sum_q \frac{\WL_0}{\theta'(qb)}
\end{align}

However, if one processes the audio stream block wise, the worst case cannot arise for all samples in the block at once, because two worst case scenarios have a distance of at least $M=\max_q{d_q}$ (actually $M=\text{lcm}(\set{d_q})$), the next $M$ samples have for sure lower computational cost. Furthermore, the next $m=\min_q{d_q}$ samples have zero computational cost. Therefore the average costs are a suitable measure for the analysis process.

\subsubsection{Synthesis Part}
The complexity of the synthesis part is the same as that of the analysis part. Furthermore, in the synthesis part the worst case scenario can be avoided, because only the parts of the next frame buffer have to be computed in real-time, the rest can be computed later. Nevertheless, our given implementation is not optimised in this direction.

\subsubsection{Memory Costs}
Our algorithm for precomputing the windows needs $2 q\WL_0$ cells. With slight modifications it only needs $\WL_0+\sum_q \WL_q$. The smallest possible number of cells being needed is $\sum_q\WL_q$. The analysis and synthesis algorithm both need at least a buffer of size 
$\textrm{\emph{audiobuffer}}+\WL_{\max}$, where $\WL_{\max}$ denotes the window length of the longest windows used in the analysis and synthesis, and \emph{audiobuffer} the buffer length in which the audio is processed. 

The frame elements for our tests below needed between 50 and 400~MB.

\section{Computational Error}
\label{sec:error_estimates}

The measured $err$-numbers are the difference between the averaged RMS-amplitude in dB of the input signal and the analysed-synthesised output signal (i.e. negative signal to noise ratio). For comparison: 16\,bit quantization (which is CD-quality) has $err\simeq-96$\,dB, 8\,bit quantization has $err\simeq-50$\,dB. 
The tests were conducted with the PD-patch available at \href{http://tommsch.com/science.php}{tommsch.com/science.php} 
in real-time over a time of about 20\,s with 
double precision floating point numbers and a sample rate of $44.1$\,kHz.
We used the following stationary and non-stationary test signals

\begin{compactitem}	
	\item {\bf white:} White noise
	\item {\bf sine X:} A pure sine tone with X\,Hz
	\item {\bf const:} A constant signal
	\item {\bf clicks:} Clicks with a spacing of 1\,s
	\item {\bf beet:} Beethoven - Piano Sonata op\,31.2, length 25\,s.
	\item {\bf speech:} A man counting from one to twenty, length 20\,s.
	\item {\bf fire:} A firework, length 23\,s.
	\item {\bf atom:} A sample of sparse synthesised warped Gabor atoms which were also used for that specific test run.
\end{compactitem}

Since our method would lead to perfect reconstruction if the windows were time-shift invariant, the behaviour of the algorithm for stationary signals over a long time is of greatest interest. The constant signal is of interest since it is the one with the lowest possible frequency and our algorithm may bear problems with low frequencies due to the necessary cutting. The clicks represent the other extreme point of signals.
The atoms are of interest since they shows whether our algorithm has the ability to reproduce its atoms with high quality.

\begin{figure}[t]
	\centering
	\includegraphics[width=0.9\columnwidth]{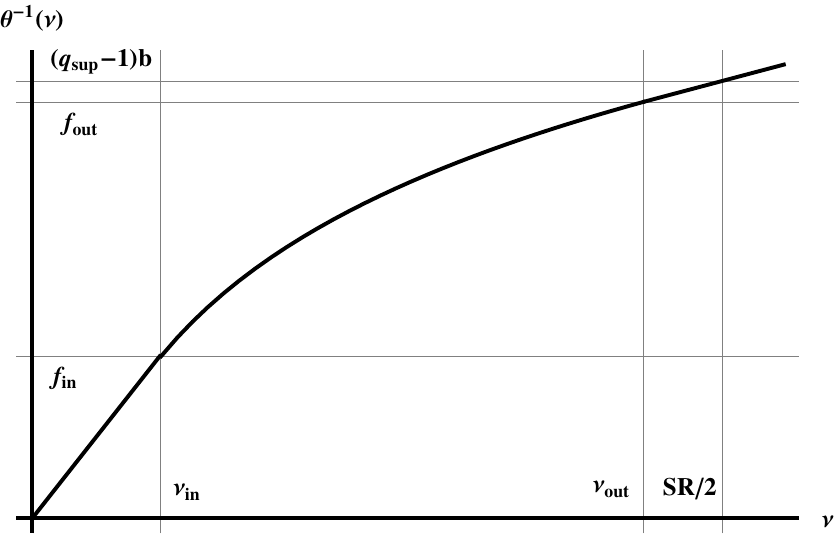}
	\caption{The used warping map $\theta$. It is $C^1$, linear between $\pm f_{in}$ and outside of $\pm f_{out}$ and passes through $SR/2$.}
		\label{fig:theta}
\end{figure}

For the warping map, we used a function which is exponentially increasing between two frequencies $f_{in}$ and $f_{out}$, namely
$\theta(f)=f_0 2^{-f/k}$ where $f_0, k\in\RR$ are constant parameters which can be set inside the PD-patch. For all the  tests we set $f_0=12$ and $k=36$. Outside of $\pm f_{out}$ and between $\pm f_{in}$ the map is linear. The function attains exactly Nyquist frequency, i.e. at $\theta((q_{\sup}-1)b)=\SR/2$. The frequencies $f_{in}$, $f_{out}$ are chosen such, that the resulting map is $C^1$. 
See Figure~\ref{fig:theta} for a plot of $\theta^{-1}$.

\subsection{Raised Cosine Window}

\begin{figure}[t]
	\centering
	\includegraphics[width=0.5\columnwidth]{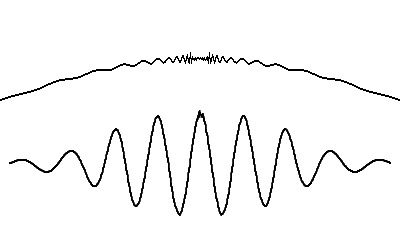}
	\caption{The picture shows a warped raised cosine window in the time domain. The upper part is a magnification around zero. Ripples, due to the slow decay in the frequency domain, are clearly visible.
		\label{fig:ripples}}
\end{figure}
As proposed in~\cite{Eva14} we first performed tests with a raised cosine window.
\begin{align}
h(t)= \begin{cases} 
\sqrt{\frac{2b}{R}} \cos \frac{\pi t}{\WL} & -\frac{\WL}{2}\leq t \leq+\frac{\WL}{2}\\
0 & \text{otherwise}
\end{cases}
\end{align}
with Fourier-Transform
\begin{align}
\hat{h}(\nu)=
\sqrt{\frac{b}{2R}} 
\Big(\sinc(\nu \WL-\tfrac{1}{2})+\sinc(\nu \WL+\tfrac{1}{2})\Big)\end{align}
where $\WL$ is the total duration of the window, $R\geq 3$ is an integer, $f_0=\frac{3}{2\WL}$, $a=\frac{3}{2R f_0}$, $b= \frac{2f_0}{R}$. See~\cite[Section 4]{Eva14} for an ansatz about how to compute these parameters. This window has a very slow decay in the frequency domain after warping:
\begin{align}
\hat{\ttphi}_{0,q}(f)=\sqrt{\frac{d_q}{a}}\hat{h}(\theta_e^{-1}(f)-qb)\simeq \frac{1}{\log_2 f}
\end{align}
This means that either the warped windows are not bounded anymore (i.e.: $\ttphi_q\notin L^\infty(\RR)$) or that they are discontinuous, which is clearly visible in the warped windows, a plot of one window can be seen in Figure~\ref{fig:ripples}. 
Hence the computation of the warped windows with the IDFT bears numerical errors. Also the test results with this window were suboptimal, yielding an error between \mbox{$-50$\,dB} and \mbox{$-60$\,dB}, depending on the chosen parameters and input signal.

Changing the parameter $R$ or the parameter $C_d$ had a significant effect on the error. In Figure~\ref{results:influence_R} on can see the influence of $R$. This can be expected, since the essential bandwidth, in which the phase is linearised, depends on that parameter. Since oversampling is directly proportional to the computational load (as computed in ~\eqref{equ:navg}), for real-time applications there is a natural upper bound. $C_d\approx 2,5$  provided good results. Too small and too big numbers for $C_d$ both gave worst results.

\begin{figure}[!ht]
	\centering
		\begin{tabular}{  c  c c r |  r }
			$4\sum \WL_q/d_q$ &	$R$& $q_{\sup}$ &  $b$~~\,&$err$  \\
			\toprule
			13.1k	& 2,0 	& 67	& 6\,Hz		& $\cgrn{-74.4}$\\
			22.1k	& 2,5 	& 83 	& 4,8\,Hz	& $\cgrn{-86.4}$\\
			25.9k	& 3,0 	& 99 	& 4\,Hz		& $\cgrn{-94.4}$
		\end{tabular}
	\caption{Influence of the parameter $R$ on the error.
		Gaussian Window, $\BW$=24\,Hz, $R$, $q_{\sup}$, $b$=variable, $\WL$=0.167\,s, $C_d=2$, \WL$_{\max}$=0.629\,s, $C_{\WL_c}$=2, $a$=0.04167\,s,  $C_{\cut}=20$, test signal: white noise. Values in\,dB. 
	}		
	\label{results:influence_R}
\end{figure}

On the contrary, changing the parameter $b$ while preserving the oversampling factor $R$ and the parameter $a$ had no effect, as long $C_b\leq 2$.

\begin{figure}[!ht]
	\centering
	\includegraphics[width=0.7\columnwidth]{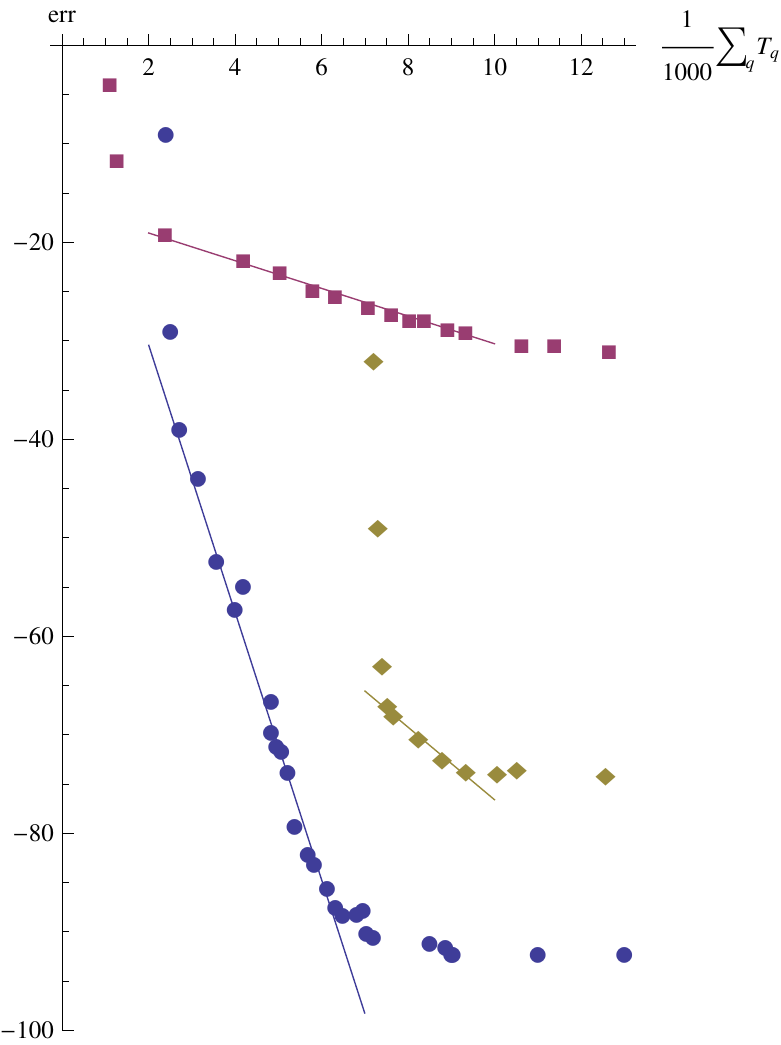}
	\caption{Influence of the truncation length of the computed window. Test signal: white noise. Red squares and yellow diamonds: Raised Cosine Window with different parameters, blue dots: Gaussian window. $C_{\cut}$=variable, all other parameters are fixed. The value of $\sum\WL_q$ is directly proportional to the error in a certain range.}
	\label{results:sum_WL_q}
\end{figure}

 The relation between the window length after cutting and the error was nearly proportional to the value of $\sum_q \WL_q$ in a certain range, see Figure~\ref{results:sum_WL_q}. From this observation one can determine a suitable cutting parameter $C_{\cut}$.
Changing the parameter $\WL_{\max}$ has clearly only an influence on the $err$ for low frequencies.

\begin{figure}[!ht]
	\centering
		\begin{tabular}{  r c  c    }
			$\nicefrac{\WL_c}{\WL_h}$ & Raised Cosine & Gaussian\\
			\toprule
			1,0 & $-63,1$ 			& $-62,8$ 		 \\
			1,2 & $-66,9$ 			& $-66,1$ 		 \\
			1,4 & $\cgrn{-70,2}$	& $\cgrn{-71,3}$ 	\\
			1,6 & $\cgrn{-71,0}$	& $\cgrn{-73,3}$ 	\\
			1,8 & $\cgrn{-71,1}$	& $\cgrn{-76,0}$ 	\\
			2,0 & $\cgrn{-71,1}$	& $\cgrn{-77,7}$ 	\\
			2,5 & $\cgrn{-71,1}$	& $\cgrn{-78,3}$ 	\\
			5,0 & $\cgrn{-71,0}$	& $\cgrn{-78,3}$ 	\\
			16,0& $\cgrn{-71,0}$	& $\cgrn{-71,5}$ 	
		\end{tabular}
	\caption{Influence of the parameter $T_c$ on the error for the Gaussian and the raised cosine window. Parameters as in Figure~\ref{results:gaussian} and~\ref{results:rcw}. Only the Parameter $C_{T_c}$ was changed. Signal: White noise. Values in dB}		
	\label{results:T_c}
\end{figure}

The parameter $C_{\WL_c}$ had no big influence as long it was about $\WL_c$ was about twice the window length $\WL_h$, see Figure~\ref{results:T_c}

The warping map has a very big influence on the error. At points where the map is not smooth the error for these frequencies is order of magnitudes higher. This can be seen in Figures~\ref{results:rcw} and~\ref{results:gaussian} for the sine with 30\,Hz signal. At this point, our used warping map is only $C^1$. For a $C^0$ warping map the error was again 10\,dB higher.

The table in Figure~\ref{results:rcw} shows selected numerical results with well chosen parameters for the raised cosine window. All values in dB, values above $-60$\,dB and below $-70$\,dB are coloured. The number $4\sum \WL_q/d_q$ denotes the average computational average complexity (see above).

\begin{figure}[!ht]
	\centering
\begin{tabular}{  l  r | l  r}
signal		& $err$ & signal & $err$ \\
\toprule
white 		& $\cgrn{-70,6}$	& clicks 	& $\cred{-57,0}$\\
sine 30 	& $\cred{-53,3}$	& beet		& $-66,8$\\
sine 440	& $-63,3$			& speech 	& $-61,9$\\
sine 20k 	& $\cgrn{-76,9}$	& fire 		& $-62,8$\\
const 		& $\cgrn{-84,7}$	& atom		& $-62,7$
\end{tabular}
\caption{Test results for the raised cosine window. $\BW$=24\,Hz, $R$=7, $q_{\sup}$=229, $\WL$=0.30\,s, $\WL_c$=1.4\,s, $\WL_{\max}$=0.5\,s, $a$=0.04167\,s, $b$=1.714\,Hz, $C_b$=2, $C_d$=4, $C_{\cut}$=55, $4\sum \WL_q/d_q=33.6$k. Values in dB.}
\label{results:rcw}
\end{figure}

\subsection{Gaussian Window}
In order to obtain a window with proper decay in the frequency domain after warping, we used a Gaussian window. This window does not overlap-add to one. Hence higher overlap is necessary to minimize the deviation from one. The warped windows were still fast decaying in the time domain, resulting in the possibility to cut them much shorter then the warped raised-cosine windows which compensated the high computational load due to the high overlap.

Our used Gaussian window and its Fourier Transform is
\begin{align}
h(t)=C\sqrt{\frac{b}{2R}} e^{-\frac{1}{4} \frac{t^2}{\WL^2}}, \quad \hat{h}(t)=C\,\WL\sqrt{\frac{b}{R}}e^{-t^2\WL^2}
\end{align}
where $\WL$ is the total duration of the window, $R\geq 2$ the overlap factor is an integer, $f_0=\frac{R}{2\WL}$, $a=\frac{\WL}{R}$, $b= \frac{1}{aR}$ and
$C\simeq 0.893249\cdots$ is a constant factor used to approximate the overlap-add to one condition.

\begin{figure}[!ht]
	\centering
	\begin{tabular}{  l  r | l  r}
		signal		& $err$ & signal & $err$ \\
		\toprule
		white 		& $\cgrn{-71,3}$	& clicks 	& $\cred{-57,3}$\\
		sine 30		& $-61,5$			& beet		& $\cgrn{-70,2}$\\
		sine 440 	& $\cgrn{-70,0}$	& speech 	& $-69,8$\\
		sine 20k 	& $\cgrn{-81,2}$	& fire 		& $-68,4$\\
		const 		& $-63,7$			& atom		& $-69,1$
	\end{tabular}
	\caption{Test results for the Gaussian window. $\BW$=24\,Hz, $R$=4, $q_{\sup}$=132, $\WL$=0.167\,s, $\WL_c$=0.8\,s, $\WL_{\max}$=0.4\,s, $a$=0.04167\,s, $b$=3\,Hz, $C_b$=2, $C_d$=2, $C_{\cut}$=1000, $4\sum \WL_q/d_q=7.3$k. Values in dB.}
	\label{results:gaussian}
\end{figure}

This window has a fast enough decay to ensure that the warped windows in the frequency domain still are in $L^1(\RR)$ and hence their Inverse Fourier transformed (i.e. the warped windows in the time domain) are bounded and continuous.
This window leads to significantly better results down to $-96$\,dB which is the limit when using PD's single precision numbers. The influence of the parameters on the error was the same as for the raised-cosine window.

The tables in Figure~\ref{results:gaussian} show selected numerical results.

\subsection{Comparison of these two windows}
 For the raised cosine a high number of bands must be used to achieve a small error. In Figure~\ref{results:sum_WL_q} the tests of the red dots are conducted using 92~bands, the tests with yellow dots were with 229~bands. Since the windows have a bad decay in the frequency domain, a high essential bandwidth has to be used too, which entails big overlap in time. and hence a very high average computational load, in our example $33.6$k~floating point operations per sample.
 If one uses similar parameters to the ones used for the Gaussian window in Figure~\ref{results:gaussian}, the error is in the range of $-36$\,dB.
 
Gaussian windows on the other hand do not overlap add to one and hence are not dual to themselves. Hence we at first used a very high overlap to minimize the deviation from 1, which turned out to be not necessary later. The fast decay in the time domain allows to cut the windows much shorter than the raised cosine window. This decreases the computational load, in our example only 7.3k flops per sample in average. In Figure~\ref{results:sum_WL_q} one can see that for the Gaussian window, with proper parameters, the error is in the range of $-90$\,dB.

\section{Conclusions}\label{sec:conclusions}
We have introduced a novel, flexible, easy way to construct frames starting from a classical Gabor frame. Tests show, that even in the \emph{painful case}, where perfect time realignment of the components is not guaranteed and hence our method does not lead to perfect reconstruction, the error can be made as small as the accuracy of single precision floating point numbers for a wide range of signals. This does not prove that the error is small for all signals, but gives a good estimation of the error to be expected. The transform is working in real time.

We are going to implement maps that can be arbitrarily defined, e.g., by means of interpolation of a selected number of points.
We will also implement Gabor Multipliers~\cite{Feichtinger2003} in the redressed warped Gabor expansion (partially done already with the external \emph{winary\~}). 

Since we already implemented this method as a Pure Data external, it is ready to use for audio-applications. The whole external explained in detail as well as the source code can be found online at~ \href{http://tommsch.com/science.php}{tommsch.com/science.php} and in~\cite{MEJ15}.

\section{Acknowledgements}
Many thanks to the great number of suggestions by the great number of anonymous reviewers on how to improve the paper.

\nocite{*}
\bibliographystyle{IEEEbib}

\end{document}